# Progression and Challenges of IoT in Healthcare: A Short Review

S.M. Atikur Rahman
Department of Industrial, Manufacturing and Systems Engineering
University of Texas at El Paso
El Paso, TX 79968, USA

Sifat Ibtisum
Department of Computer Science
Missouri University of Science and Technology,
Rolla, Missouri

Priya Podder
Dhaka National Medical College
Dhaka 1100, Bangladesh

S.M. Saokat Hossain
Department of CSE
Jahangirnagar University
Savar 1342, Bangladesh

## ABSTRACT
Smart healthcare, an integral element of connected living, plays a pivotal role in fulfilling a fundamental human need. The burgeoning field of smart healthcare is poised to generate substantial revenue in the foreseeable future. Its multifaceted framework encompasses vital components such as the Internet of Things (IoT), medical sensors, artificial intelligence (AI), edge and cloud computing, as well as next-generation wireless communication technologies. Many research papers discuss smart healthcare and healthcare more broadly. Numerous nations have strategically deployed the Internet of Medical Things (IoMT) alongside other measures to combat the propagation of COVID-19. This combined effort has not only enhanced the safety of frontline healthcare workers but has also augmented the overall efficacy in managing the pandemic, subsequently reducing its impact on human lives and mortality rates. Remarkable strides have been made in both applications and technology within the IoMT domain. However, it is imperative to acknowledge that this technological advancement has introduced certain challenges, particularly in the realm of security. The rapid and extensive adoption of IoMT worldwide has magnified issues related to security and privacy. These encompass a spectrum of concerns, ranging from replay attacks, man-in-the-middle attacks, impersonation, privileged insider threats, remote hijacking, password guessing, and denial of service (DoS) attacks, to malware incursions. In this comprehensive review, we undertake a comparative analysis of existing strategies designed for the detection and prevention of malware in IoT environments.

## 1. INTRODUCTION
The role played by information systems has grown significantly, particularly within the healthcare industry [1]. From electronic health records to cloud-based systems, information technology (IT) has consistently benefited the healthcare sector. As IT continues to advance, information systems have emerged as pivotal tools for enhancing healthcare and its management [2]. Consequently, there is a growing emphasis on developing new business models aimed at facilitating access to healthcare services for all stakeholders through improved information dissemination [3].

One of the most recent applications of IT in healthcare is the Internet of Things (IoT). The IoT can be defined as the interconnection of intelligent objects or devices via the Internet, leading to novel applications and innovative services [4]. These devices, equipped with extensive human detection capabilities, find utility in the medical field for remote health monitoring, early diagnosis, and elderly care [5, 6]. IoT applications in healthcare have the potential to enhance patient well-being and reduce service costs, for instance, by preventing unnecessary hospitalizations and ensuring better care for individuals in critical conditions. IoT-based healthcare services, spanning the entire value chain, are poised to revolutionize the healthcare sector [7].

Over recent years, substantial research has been conducted on IoT projects within the healthcare domain. IoT research, especially in relation to innovative healthcare applications, has witnessed rapid growth [8, 9]. For instance, Islam et al. [10] conducted an analysis of IoT-based industrial applications and trends in healthcare. Their study scrutinized various IoT and eHealth policies and regulations. Gundala et al. conducted a comprehensive review, identifying critical factors that enable a deeper understanding of the potential and challenges of implementing IoT applications in healthcare [11]. Bhuiyan et al. presented a review that classified existing IoT-based healthcare networks and provided an overview of prospective networks [12]. In this context, they analyzed IoT healthcare protocols and engaged in an extensive discussion. Furthermore, they initiated a comprehensive survey of IoT healthcare applications and services, offering profound insights into IoT healthcare security, its requisites, challenges, and privacy concerns within the healthcare IoT landscape [12].

## 2. IOMT AND MEDICAL SIGNALS
In [13], a research study employed a multi-sensor platform, encompassing two-channel pressure pulse wave (PPW) signals and one-channel ECG, to estimate blood pressure (BP). From the acquired signals, they extracted 35 physiological and informative features. To mitigate dimensionality and identify the most pertinent indicators for individual subjects, a weakly supervised feature (WSF) selection method was introduced. Furthermore, they harnessed a multi-instance regression algorithm to amalgamate these features, thus enhancing the BP prediction model. Another study, delineated in [14], introduced a method for emotion recognition and classification across diverse individuals. This approach integrated significance testing and sequential backward selection with a support vector machine (ST-SBSSVM) to heighten the accuracy of emotion recognition. Gu et al. [15] put forth a real-time monitoring system designed to bolster precision and mitigate risks for mine workers. This system involved the fusion of data from multiple sensors, integrating concepts from situation awareness and the Internet of Things (IoT). A random forest (RF) SVM-based model was employed to assess situation levels and merge data. Through simulation analysis, they demonstrated an RMSE below 0.2 and a TSQ (Time Series Quality) not exceeding 1.691 after 200 iterations. In a distinct approach, [16] introduced an ensemble method with data fusion capabilities. This technique fused data sourced from a body sensor network (BSNs) and incorporated it into an ensemble classifier for the prediction of heart disease. Steenkiste et al. [17] presented a robust model aimed at enhancing the performance and reliability of sleep apnea prediction via sensor fusion. They collected and integrated multi-sensor data, including oxygen saturation, heart rate, thoracic respiratory belt, and abdominal respiratory belt, utilizing a method that incorporated backward shortcut connections. In the domain of IoT applications, an event-driven data fusion tree routing algorithm was introduced in [18]. This paper delved into health and sports information gathering systems, comprised of terminal nodes and client management systems. The algorithm was tailor-made to accommodate IoT communication characteristics and incorporated visual modeling, yielding superior accuracy and timeliness compared to alternative methods.

Chiuchisan et al. [19] outlined the design of a healthcare network for monitoring at-risk patients in smart intensive care units (ICUs) through an IoT model. They leveraged sensors and the Xbox Kinect to track patient movements and facilitate real-time alerts to physicians, thus enhancing patient care. Sharipudin and Ismail [20] proposed a healthcare monitoring system for data management in patient monitoring, integrating healthcare sensors to measure vital parameters, with data transmission to cloud storage for medical reference.

Dimitrov [21] expounded on IoMT (Internet of Medical Things) applications and Big Data in healthcare, illustrating their potential to support innovative commercial models, streamline work processes, improve customer experiences, and enhance outcomes. The usage of wearable sensors and mobile applications played a pivotal role in addressing various health needs and gathering patient data for health education.

In [22], early warning score systems were established based on vital sign characteristics, supporting health state estimation and informed critical care decision-making. The study explored diverse machine learning techniques for risk prediction based on medical data inputs. Sanyal et al. [23] introduced a federated filtering framework (FFF) for central fog server data forecasting using aggregated models from IoMT devices. This framework addressed critical concerns regarding energy efficiency, privacy, and latency for resource-constrained IoMT devices.

Finally, Luna-del Risco et al. [24] deliberated on the recognition, implementation challenges, and threats to wearable technology adoption in the Latin American healthcare system. The analysis revealed key challenges related to human resource allocation, healthcare connectivity, funding for health programs, and healthcare disparities. Wearable sensors were considered a valuable component of the solution to address healthcare challenges in the region.

## 3. ARCHITECTURE OF HEALTH IOT SYSTEM

Advancements in IoT and wearable technology are driving the adoption of personalized healthcare systems. These technologies create a connected environment, enabling various healthcare services such as remote health monitoring, fitness tracking, and nutrition programs. Figure 1 depicts the service-oriented architecture of a personalized Health IoT system, with each layer representing specific services.

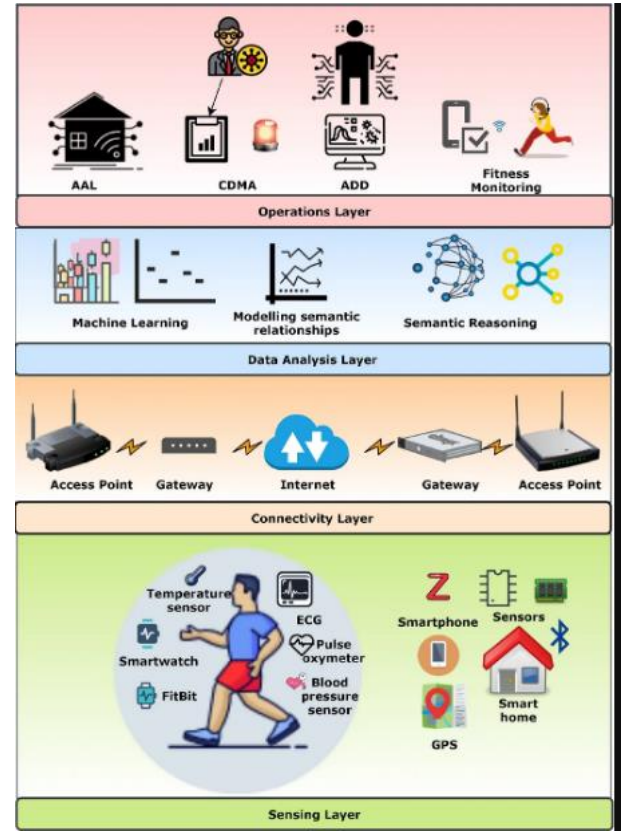

**Figure 1: Service-oriented architecture of a personalized Health IoT system (Adopted from [25])**

The Sensing layer comprises wearable sensors responsible for tracking body parameters, along with wireless communication technologies like Bluetooth and Zigbee. The Connectivity layer transfers sensor data to the Internet using devices like routers and gateways. The Data Analytics layer, positioned between the cloud and connectivity layers, ensures quality of service (QoS) requirements, reduces latency by caching data closer to users, and conducts data processing and analytics using machine learning. The topmost layer, the Operations layer, is primarily composed of cloud servers. It performs advanced processing and analysis of sensory data, delivering healthcare services like active alerts, fitness monitoring, and more.

Figure 2 illustrates the conventional three-tier architecture of an H-IoT system, comprising the Sensing layer, Network layer, and Cloud layer. The Sensing layer records vital body data using sensors and performs minimal computation before forwarding it to the Edge layer through the network. Various wireless communication protocols like Bluetooth, Wi-Fi, and Zigbee transmit data to the Edge/Fog Node via gateway devices. These nodes perform initial data processing near the Sensing end. The Cloud Layer analyzes the extensive data collected from sensing layer sensors to provide essential H-IoT applications. A tele rehabilitation system based on IoT was demonstrated in [28] the system's capabilities in real-time remote health monitoring, enhancing the quality of life for older and disabled individuals, disease prediction, preventive medication, and healthcare service management in hospitals and emergency rooms [29, 30].

Moreover, the integration of blockchain with IoT devices in the IoMT ecosystem greatly facilitates the global sharing of patient medical data. By securely sending IoMT data to the blockchain through smart contracts, the system effectively reduces data tampering and information manipulation [31-33]. Blockchain security policies establish trust among stakeholders. Utilizing blockchain's decentralized storage concept simplifies and enhances the transparency of data collection, sharing, storage, and maintenance [34, 35]. Moreover, it ensures rigorous control over patient privacy policies.

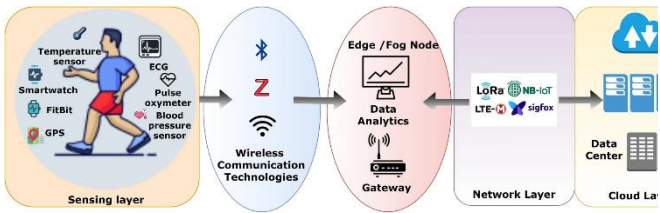

**Figure 2: The three-layer generic architecture of H-IoT system (Adopted from [25])**

## 4. ROLE OF CLOUD COMPUTING IN HEALTHCARE

Cloud computing (CC) has a positive impact on healthcare by enhancing efficiency and reducing costs. It offers advantages such as easy storage, sharing, and maintenance of medical data, as well as automated backend processes. To address the challenges of handling large datasets and providing physicians with access to patient health information, cloud infrastructure integrated with IoT devices is a viable solution [36]. However, two key concerns in a cloud based IoT system are data security and privacy, as well as the ability to support machine learning (ML) and deep learning (DL) models.

**(a) Cloud Data Storage:** This is a CC model that stores data on the Internet. Data plays a crucial role in diagnosing and analyzing diseases, making it necessary to store data in a repository (database) for research purposes. It allows hospitals and clinics to store vast amounts of patient and staff data that can be retrieved in case of emergencies. Cloud storage comes in three types: object storage, block storage, and file storage. Its advantages include efficient data maintenance and management, quick deployment, and cost-effectiveness [37].

**(b) Data Processing and Analysis:** The cloud provides various data processing techniques, including computer offloading, online data processing, electronic data processing, real-time processing, and machine learning. Among these, machine learning, specialized data mining, and offloading processing are the most relevant methods. Computational offloading involves transmitting raw data to the cloud for further processing, particularly suitable for handling complex data. Machine and deep learning techniques such as SVM, KNN, decision trees, random forests, CNN, LSTM, and autoencoders are employed for data analysis, while data mining techniques are used to extract valuable information from repositories. High-performance computing integrated with smartphone processing allows for the execution of sophisticated algorithms, resulting in accurate outcomes and extended smartphone service life due to reduced internal computations.

**(c) Data Cleaning:** Noise, or significant modifications in image pixel values, can degrade diagnosis performance, particularly when noise is introduced during data acquisition and transmission. Data cleaning is essential, and filters such as mean filters, median filters, Gaussian filters, and Wiener filters can be applied for denoising. Image filtering replaces pixel values with the average values of neighboring pixels, resulting in smoother images. Figure 3 describes the advantages of the cloud in the healthcare system. Characteristics of cloud based IoT systems are as follows:

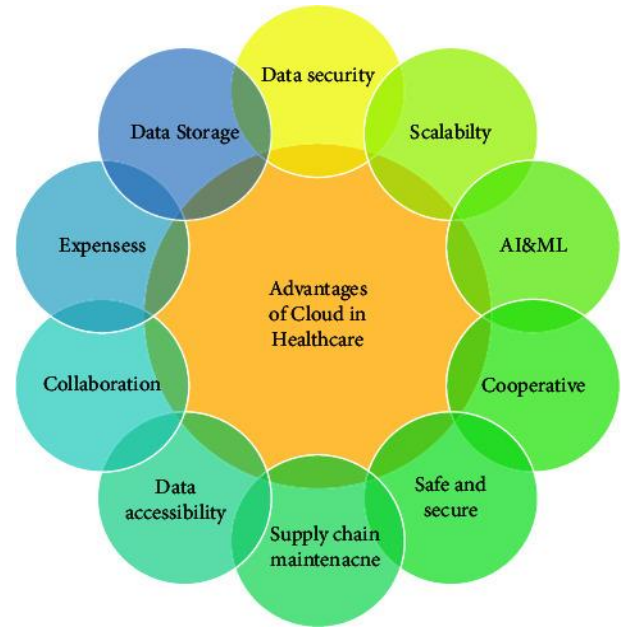

**Figure 3: Cloud Computing in Healthcare (Adopted from [38])**

## 5. APPLICATIONS OF CC AND IOMT IN DISEASE DETECTION

### 5.1 IoMT in Breast Cancer Detection

A significant volume of research has been conducted in the field of breast cancer detection and diagnosis, with some notable studies outlined here. Spanhol et al. [39] developed a system employing a variant of Convolutional Neural Network (CNN) known as ALexNet to classify malignant and benign cases using histopathological images. In [40], Ertosun and Rubin introduced a model to identify the presence of masses in mammography images and subsequently locate these masses within the images. In [41], a deep learning model was designed to detect mitosis, utilizing feature extraction techniques like CNN. The output from this was fed into a Support Vector Machine (SVM) classifier to identify breast tumors. For mitosis detection in breast histopathology scans, a deep contour-aware network was introduced [42]. Xu et al. [43] proposed a classification model using a Stacked Sparse Autoencoder (SSAE) to classify nuclei in histopathological breast images. A greedy algorithm was applied to optimize the SSAE. Additionally, breast cancer detection was performed using a mammography image dataset. In [44, 45], a hybrid model combining CNN and SVM was developed to detect masses in mammographic scans. CNN was trained on mammogram abnormalities, and the fully connected layer was used to extract high-level features, which were then employed to train the SVM classifier. Kim et al. [46] suggested a 3D multi-view approach using CNN to analyze bilateral features via Digital Breast Tomosynthesis (DBT). Volume of Interest (VOI) was extracted from the input volume, and two different CNN algorithms were employed to extract features from the VOI. Yu et al. [47] proposed a deep learning-enabled system for breast cancer diagnosis in remote healthcare settings using 5G technology. This approach involves three steps: acquiring input data from hospitals via 5G, applying a transfer learning algorithm to develop a diagnostic system, and deploying this model on an edge server for remote diagnosis. Modified VGG (MVGG) was proposed in [26] to detect breast cancer from 2D and 3D images of mammograms.

## 5.2 IoMT in Diabetes Detection

Numerous methods and techniques have been proposed for the detection of diabetes. In [48], various machine learning techniques, including SVM, decision trees, and Naive Bayes, were applied.

In [49], a hybrid approach utilizing Principal Component Analysis (PCA) and the Adaptive Neuro-Fuzzy Interface System (ANFIS) was employed. The development of ResNet to tackle the vanishing gradient problem is discussed in [50]. In [51], an Artificial Neural Network (ANN) architecture was introduced for diabetes prediction. This model aimed to minimize the error rate during training and achieved an 87% accuracy rate with an estimated error rate of 0.01%. In [52] a computer-aided system was presented for the detection of diabetic retinopathy using digital signals obtained from retinal images. The primary objective was to classify nonproliferative diabetic retinopathy from retinal images. A Convolutional Neural Network (CNN) approach was introduced in [53] for segmenting blood vessels in diabetic patients, followed by classification and the extraction of discriminant patterns. [54] developed a shallow CNN framework for this purpose. Finally, in [55], a residual network (ResNet) was developed for the automatic classification of retinopathy images.

## 6. MALWARE THREATS IN IOMT

Security concerns related to IoT devices are on the rise due to the rapid development and widespread deployment of IoT systems. This trend has created opportunities for various types of cyberattacks within the IoT environment, facilitated by internet connectivity. The situation becomes particularly critical in the context of the Internet of Medical Things (IoMT), which involves communication and control of smart medical devices [56-59]. For instance, if an attacker gains remote control over a smart medical device, they can pose a serious threat to the patient's life, such as causing a smart pacemaker to deliver a potentially lethal shock.

The emergence of various types of IoT malware compounds these security challenges. These evolving malware strains have the potential to compromise the communication within IoMT systems and manipulate smart medical devices. In the IoMT context, malfunctions of smart medical devices may also occur, such as the unintended release of insulin from an implanted blood glucose monitoring system [60].

Several categories of malware are pertinent to this discussion [61]-[64]:

**(a) Spyware:** This type of malware operates by covertly monitoring a user's activities without their consent. It engages in malicious activities like collecting keystrokes, tracking user behavior, and harvesting data, including account credentials and financial information like credit card numbers. Spyware often exploits software vulnerabilities and can attach itself to legitimate programs.

**(b) Keylogger:** Keyloggers are malicious code designed to record a user's keystrokes. This means that everything a user types, including login information and passwords, can be captured. Keylogger attacks are particularly potent and can bypass strong password protection. To mitigate this threat, it is recommended to implement multi-factor authentication (MFA), which combines username, password, smart card, and biometric data.

**(c) Trojan Horse:** This malware disguises itself as a legitimate program to deceive users into downloading and installing it. Once installed, it provides unauthorized remote access to the infected system, enabling hackers to steal sensitive data like account and credit card numbers. Trojan Horses can also introduce additional malicious software into the system for various illicit purposes.

**(d) Virus:** Viruses are malicious programs capable of replicating themselves and spreading to other systems. They achieve this by attaching themselves to different programs, activating their code when a user launches one of the infected programs. Viruses can be employed to steal information, damage host systems, and create botnets.

**(e) Worm:** Worms propagate across networks by exploiting vulnerabilities in operating systems. They can disrupt host networks through excessive bandwidth usage and server overloads. Some worms carry payloads designed to harm host systems. Hackers frequently use worms to steal sensitive data, delete files, or establish botnets.

These security threats underscore the importance of robust security measures and continuous vigilance in the IoMT and broader IoT ecosystems.

## 7. FUTURE DIRECTIONS

The Internet of Things (IoT) is rapidly transforming the healthcare industry, offering promising future directions for progression [65-73].

(a) **Remote Patient Monitoring:** IoT devices enable real-time monitoring of patients' vital signs, chronic conditions, and overall health. This can significantly improve the quality of care and reduce the need for frequent in-person visits, particularly in the context of an aging population.

**(b) Enhanced Data Analytics**: IoT-generated data allows for advanced analytics and predictive modeling. Machine learning and AI can identify trends and predict potential health issues, aiding in early intervention and personalized treatment plans.

**(c) Wearable Technology:** Wearable IoT devices, like smartwatches and fitness trackers, are becoming increasingly sophisticated. They offer continuous health tracking, including ECG, blood pressure, and sleep patterns, empowering individuals to take control of their health.

**(d) Telemedicine:** The integration of IoT in telemedicine enables secure, real-time video consultations and the remote exchange of patient data. This is especially valuable in providing healthcare access to underserved or remote areas.

**(e) Drug Adherence Monitoring:** IoT can improve medication adherence by providing reminders and tracking dosages. This is crucial for chronic disease management.

## 8. CONCLUSION
Smart healthcare is a well-explored field encompassing IoT, IoMT, medical data, AI, edge and cloud computing, each with its own strategies and speeds. This tutorial paper aims to categorize and compare IoT, IoMT, AI, cloud computing, and malware threats in smart healthcare. Traditional AI-based healthcare systems might not be readily accepted by medical professionals. Hence, employing explainable AI systems allows doctors to visualize disease detection or classification. Edge resources can be optimized effectively using intelligent algorithms. Furthermore, the evolution of next-gen wireless networks, including 5G and beyond, accelerates global healthcare accessibility. Additionally, federated deep learning and edge computing are simplified and strengthened.